\RequirePackage{fix-cm}
\documentclass[twocolumn]{article}

\usepackage[a4paper, margin=0.5in]{geometry}
\usepackage{graphicx}
\usepackage{graphicx}
\usepackage{calc}
\usepackage[table]{xcolor}
\usepackage{authblk}

\def\keywords#1{\\\\\textbf{Keywords}. #1.}

\begin{document}

\title{An Analysis of Twitter Users From The Perspective of Their Behavior, Language, Region and Development Indices - A Study of 80 Million Tweets}

\author{Shahab Saquib Sohail}
\author{Mohammad Muzammil Khan\footnote{Contact: mmkhan.sch@jamiahamdard.ac.in}}
\author{M. Afshar Alam}
\affil{Department of Computer Science and Engineering\\School of Engineering Sciences and Technology\\Jamia Hamdard, India}

\date{}

\maketitle

\begin{abstract}
    The need for a comprehensive study to explore various aspects of online social media has been instigated by many researchers. This paper gives an insight into the social platform, Twitter. In this present work, we have illustrated stepwise procedure for crawling the data and discuss the key issues related to extracting associated features that can be useful in Twitter-related research while crawling these data from Application Programming Interfaces (APIs). Further, the data that comprises of over 86 million tweets have been analysed from various perspective including the most used languages, most frequent words, most frequent users, countries with most and least tweets and re-tweets, etc. The analysis reveals that the users’ data associated with Twitter has a high affinity for researches in the various domain that includes politics, social science, economics, and linguistics, etc. In addition, the relation between Twitter users of a country and its human development index has been identified. It is observed that countries with very high human development indices have a relatively higher number of tweets compared to low human development indices countries. It is envisaged that the present study shall open many doors of researches in information processing and data science.
\end{abstract}

\section{Introduction}
The popularity of social networking sites has increased tremendously in recent years [1]. The number of Twitter users were reported approximately 40 million in 2009 [2] which has risen to 330 million in 2019 [3]. Twitter users used to express their emotions and views through 140 characters (which has now extended to 280) texts or they can share images and videos [4] termed as ‘tweet’. Since the average tweets have gained a rapid climb and according to the current study, 6k tweets per second are tweeted that implies $5*10^8$ (five hundred million) tweets per day [5]. In a recent development, it is reported that Twitter has a potential social influence for online campaigning [6] and sharing the current events and spreading breaking news [7]. The proliferation of online social networking has also opened various research dimensions and hence attracted researchers to explore the possibilities. A number of researchers have worked on Twitter data to predict stock market [8, 9], disaster management [10] and to analyse sentiments of the users through their tweets [11–15]. Some of them have tried to identify the relation between online news and Twitter [16–18] and few of the researchers have also raised the concern over privacy and security issues [19, 20]. However, the overall picture of the users and their relations with respect to the geographical region they belong and the political status of their place of living would be very interesting and useful. It would surely give an insight into the users over Twitter and estimate the real influence this social networking platform may have. Recently, the authors have argued the need and importance of comprehensive research to explore increasing social network-based information and its effect [21]. Therefore, it is of great importance and common interest of the masses to understand the Twitter trends viz. usage pattern, users’ distribution geographically and linguistically, the most frequent users and the countries with the most tweets. 
Here, we have crawled approximately 86.79 million of data (tweet objects) from Twitter with the help of the Application Programming Interface (APIs). In addition to the tweets, their related information like user ID, location, language, time, whether it is the original tweets or re-tweets (re-tweets are forwarded tweets of others from some one’s profile), etc. have been classified and saved. We have used statistical measures to identify the various trends and behaviour over Twitter. On the one hand, the present work adds value to the existing literature and on the other, it enables readers to identify the trends in twitter and analyse it from various interesting perspectives. Our major contributions in this article can be summarized as follows – 

\begin{enumerate}
    \item A detailed procedure to crawl the Twitter data has been discussed.
    \item A new algorithm to detect the location of the users (which is ambiguous in crawled data) have been suggested.
    \item An analysis of the users, their geographical locations and comparison of the country with respect to their Twitter users have been made.
    \item The relation between the countries human development index (HDI) and their Twitter users have been explored and discussed.
    \item The users with most Tweets, most frequent words which have been used and the most frequently used languages have been identified and discussed.
\end{enumerate}

The rest of the paper has been organized as follows. Section 2 describes the background of the work and section 3 discusses the data description, the methodology involved in data gathering and algorithm to detect users’ locations. In section 4, the parameters for the experiments have been discussed whereas section 5 deals with the detail results obtained in this work. All these results are discussed in section 6. We conclude in section 7 which also opens the various future direction.

\section{Background}
Researchers are investigating the impacts and influences on everyday life from a social networking platform for quite some time now. Whether they apply their studies to find patterns of influence between users and their usage with their respected audience [22] or for what purpose does a specific platform may fulfil, i.e., to spread news or converse with their audience [23]. Researchers have also reiterated [24] and found new results from previous datasets [22] and applied their methods to extract and infer meaningful information. They identified location, gender, and race/ethnicity all using a few attributes like names and user-reported locations. The authors further investigated the demographics of users to compare them. 

Opinion mining, sentiment analysis, usage analysis, etc. have also been used to predict the stock market [25, 26], book sales [27], movie-ticket sales [28, 29], product sales [30], infections, diseases, and consumer-spending [31]. Authors have also incorporated Twitter data for healthcare purpose and disease prediction [32, 33]. Researchers have proposed a new generic-type of recommendation filtering based on social media usage, which incorporated purchase intent, information extraction, and recommendation, instead of the traditional and limited approaches for the recommender systems used in e-commerce [34]. Researchers have studied the data to assess the prevalence of obesity [35], physical activities [36], and understand biasness in the crowdsourced recommendation towards trends shown, and actual trends [37]. The authors have proposed a system [38] to continuously monitor Twitter for emotional tweets to infer the current dominant emotion in cities. Some researchers have performed user profiling [39] and some detected context-depended speech patterns and dialects [40–42] to explore further research dimensions in the concerned field.

A holistic view of the users and its projection from a political and geographical point of view would open the many doors of research in the area of linguistic, social science, data science, and privacy \& security, etc. The very aspects of the users from these perspectives have not been well addressed in the literature and would benefit from understanding the users’ behaviour over Twitter. Therefore, we have incorporated these aspects in this paper, in addition, the data crawling from social network platforms to identify the users’ geographical location is also discussed.

\section{Data-set and Methodology}

\subsection{Acquiring the data}
To acquire the data for analysis we used the official Twitter Application Programming Interface (API) to collect the latest tweets. We requested the API for new tweets after every 500ms - 2000ms interval. With a rate limit of 450 requests per 15 minutes, we ran it for around 23 days continuously and iterated over the responses. Only the responses which had the location field reported and language detected were saved to the disk. If either of them, i.e. location or language, were undefined, our program skipped them. After 23 days, we had iterated over 86.79 million (86,786,849) tweets and saved 58.15 million (58,147,189) tweets. After this process, we ran a check to verify the integrity and degree of facsimile. In this test, we found that the syntax of our dataset was firm but we had around 8.2 million (8,201,949) duplicate records which were then removed from the dataset. 
\subsection{Location detection}
After verifying and having a dataset of distinct tweets, we then processed the user-reported locations to detect the real countries. To achieve this, we created and supervised an algorithm to detect countries and cities from plain-text (see algorithm below). The geographical location identifier was not able to detect all of the locations reported because some users didn’t use real countries and cities, instead, they used either fictional cities like \textit{Konoha}, \textit{Gotham City}, \textit{Hueco Mundo}, \textit{Asgard,} etc.
\begin{tabbing}
Func \= location (plaintext) : country \\
\> plaintext = normaliseDiacritics(plaintext) \\
\> For \= each entry in CountriesCities \\
\> \> If \= regexMatch (entry.possibleMatch, plaintext) \\
\> \> \> Return entry.details \\
\> Return null \\
\end{tabbing}

To counter this, we ran this algorithm twice. In the first round, we got the false negatives/positives and adjusted the algorithm manually, and then lastly, we updated the dataset with new attributes. There were 2.36 million (2,358,248) unknown locations which have been found and to improve the results, we analysed these locations and updated the dataset. We then ran a test to see if there was any improvement. As a result, more than 369K (369,954) entries and around 6 million (5,999,168) records can now be detected with the new dataset of countries and cities. The improvement is almost 12.01\% with a new detection rate with 0.68 precision. 

\subsection{The data processing}
Having prepared our dataset for analysis, we had 11 attributes to a single entity, i.e. tweet, which represented information as the time of creation, tweet ID, language code, detected country, detected city, ISO 3166-2 code of the country, location, name, username, whether the crawled text is a re-tweet or original-tweet and the tweet text itself. The data is now saved and imported into other software to gain further analysis from the data.

\begin{figure*}
  \includegraphics[width=1\textwidth]{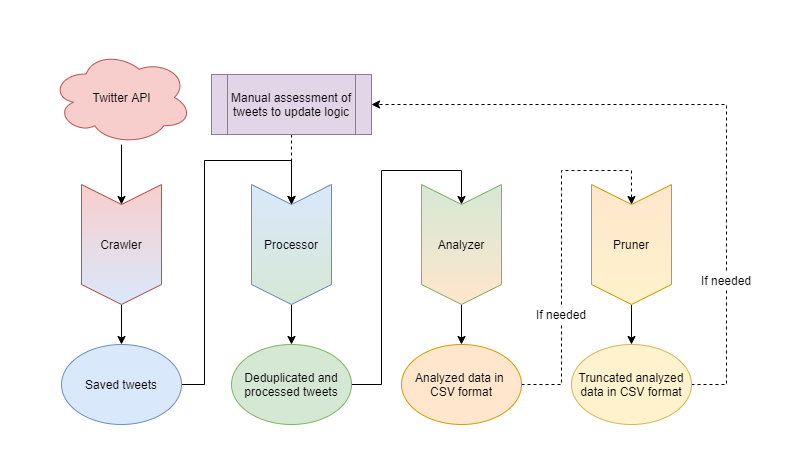}
\caption{The data crawling and processing mechanism}
\label{fig:mechanism}
\end{figure*}

Figure \ref{fig:mechanism} illustrates how the data have been crawled and further processed before it can be used to analyse further. Once we have finished the crawler after 23 days with approximately 86 million of data (tweets), the task was to remove the redundancy and noise from it. We have removed the duplicates records which were found to be around 8.2 million. Next, the users’ locations were ambiguous and tough to identify, therefore we have applied a new algorithm as stated in section 3.2 which has improved the location identification accuracy. Then we have analysed the data in .csv format and truncated whenever it was required.

\section{Experimental Analysis}

Once the data have been stored in a readable and meaningful form, we saved it as an .xlsx extension format file and accessed it to process further according to the requirements. The crawled and pruned data with the features as discussed above (tweets, ID, language etc.) is saved and accessed for observing each experiment described in the result section. Initially, we had crawled 86,786,849 tweets which were reduced to 49,945,240 after the above processes. In addition to these tweets, our investigation involved 237 countries, 12,845,715 users, and 64 different languages. The details of the tweets are saved, hence, the country from which these have been tweeted and stored and have been displayed in figure \ref{fig:distribution}. A world map [43] is used to show the distribution of tweets across the globe. Since the number of countries is larger, we have divided the countries in various ranges according to the number of tweets, and these ranges are shown through legends in the figure. Similarly, most frequent users and most frequent words, which have been stored with the tweets as their features, are shown in figure \ref{fig:500tweets}, figure \ref{fig:500retweets}, figure \ref{fig:500retweets} and figure \ref{fig:correl}. To calculate the modified spearman rank-order correlation coefficient [44], following (\ref{eq:spearman}) is used.

\begin{equation}
rs' = 1 - \frac{\sum_{i=1}^{n}(i-v_i)^2}{m([max\ j =\ _{1}^{m}\{v_j\}]^2 - 1)}
\label{eq:spearman}
\end{equation}

Where $rs'$ denotes modified spearman rank-order correlation coefficient, $V_i$ and $V_j$ are two given list (like a list of top 20 countries with very high human development indices, etc); $i$ and $j$ are the total number of elements in the list $V_i$ and $V_j$, respectively. We have calculated $rs’$ between the countries ranking as mentioned in UN human development report [45] for all the four categories viz. very high human development indices (HDI), high HDI, medium HDI and low HDI, and ranking of the corresponding countries according to the number of tweets they generate. The top 20 countries have been taken into consideration for the first three categories and least 20 ranking for the fourth category, respectively. Since the fourth category indicates low developed countries, we have taken the least from the list. The correlation of the ranking is shown through a scatter plot in figure \ref{fig:correl} and calculated $rs’$ is shown in figure \ref{fig:correl}.

\section{Results}
\begin{figure*}
  \includegraphics[width=1\textwidth]{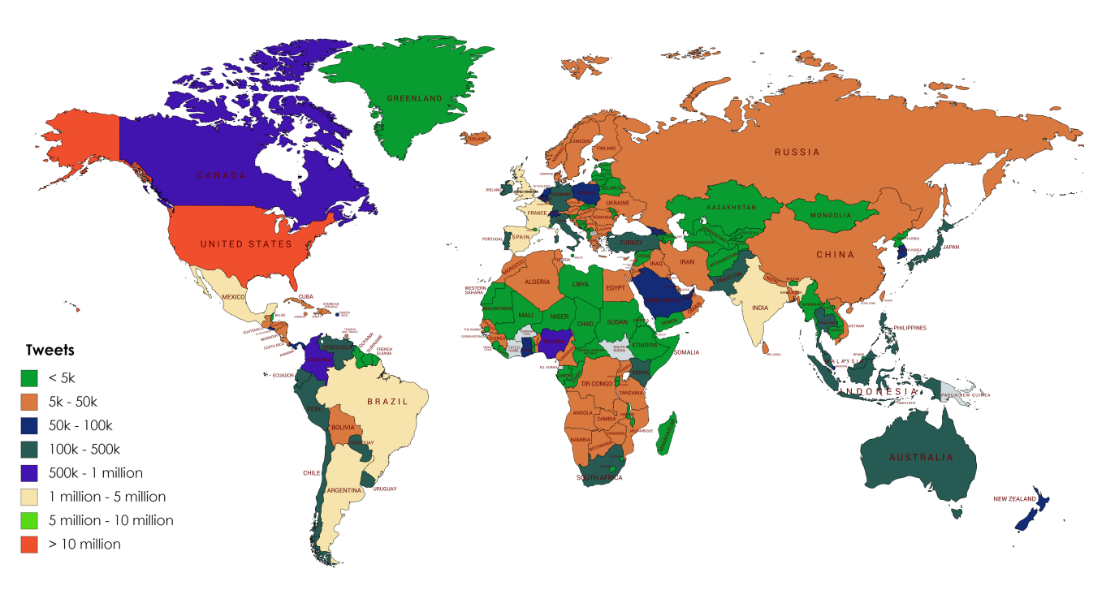}
\caption{The countries across the world and the Tweets distribution from these countries}
\label{fig:distribution}
\end{figure*}

Figure \ref{fig:distribution} shows the total tweets and the respective countries from where these have been tweeted. The countries are grouped and have been classified into eight categories on the basis of the number of tweets where total tweets are: \textit{i) less than 5000 (5k), ii) between 5k and 50k, iii) between 50k to 100k, iv) between 100k and 500k, v) between 500k and 1 million, vi) between 1 million and 5 million, vii) between 5 million and 10 million and viii) more than 10 million}. Each category is labelled with a particular colour, the legends in the figure show the group to which these countries belong. The count of tweets for most of the countries is less than 5000, i.e. number of countries having a smaller number of tweets is higher. 123 countries belong to group 1 where no country has tweets more than 5000, and the total contribution of these countries for the total tweets is 171,500. The total number of tweets for group 2 countries is 1,196,148 from 68 members. We infer here that 80\% of the world countries even do not contribute 4\% of the total tweets. The group 3 countries have 1,082,312 numbers of total tweets which are constituted with 14 countries. There are 20 countries in group 4 which contribute a total of 4,923,124 tweets. 1,884,290 tweets have been tweeted from three countries, Nigeria, Canada, and Mexico of group 5. 

The tweets from group 6 countries have been recorded to 12,650,823 from 7 members. These 7 countries contribute 33.76\% of total recorded tweets. The USA alone leads the category with 12,299,716 tweets. One thing worth to mention here is the unidentified location of the tweets for those Twitter handles who hide their location details and Twitter does not provide any mechanism to trace their real or geographical location through API. 24\% of the total recorded tweets have not been identified the place from it is operated. 

\begin{figure}
  \includegraphics[width=0.48\textwidth]{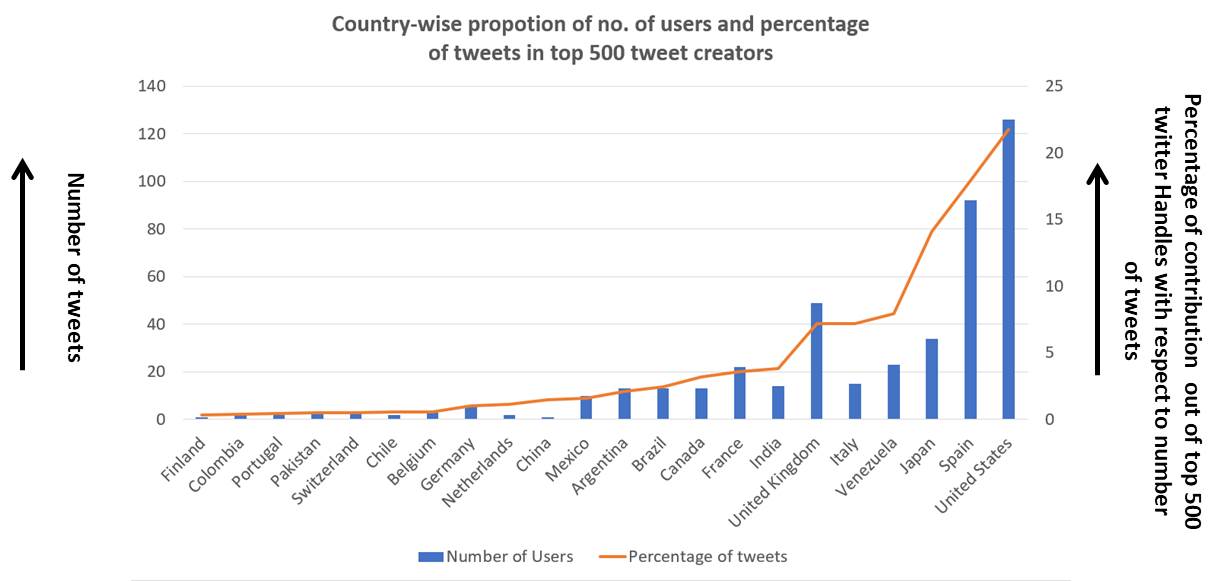}
\caption{Country-wise details of the number of Twitter Handles among top 500 Tweets generator and the percentage of these users from the respective countries}
\label{fig:500tweets}
\end{figure}

\begin{figure}
  \includegraphics[width=0.48\textwidth]{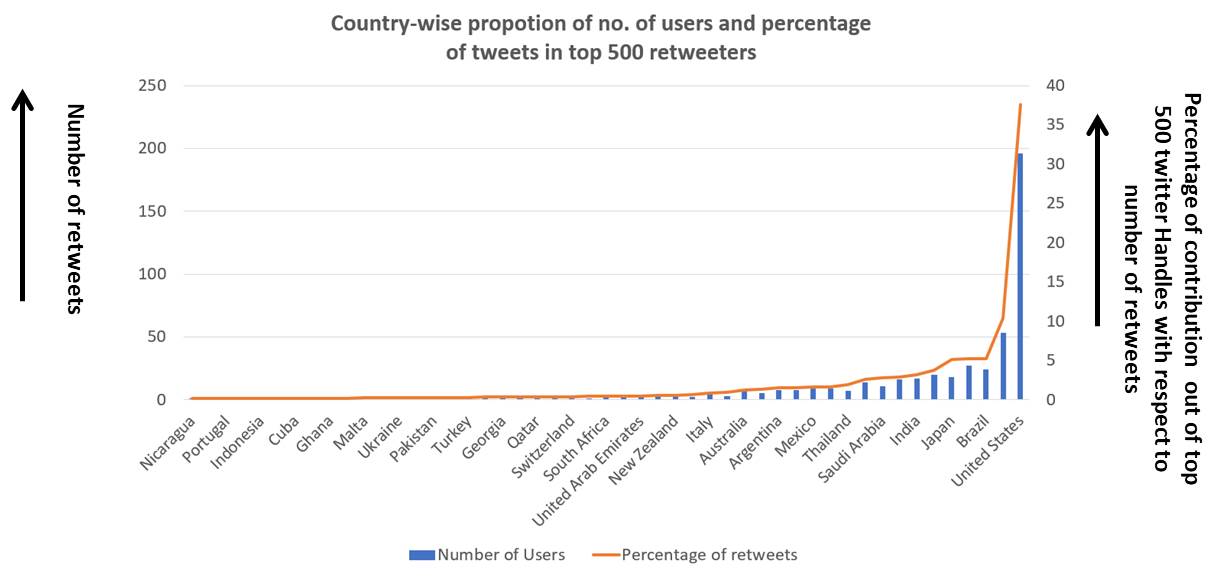}
\caption{Country-wise details of the number of Twitter Handles among top 500 Re-Tweets generator and the percentage of these users from the respective countries}
\label{fig:500retweets}
\end{figure}

In Figure \ref{fig:500tweets} and Figure \ref{fig:500retweets}, the top 500 Twitter handles who contribute most with respect to Tweets and Re-Tweets are shown, respectively. We intend here to identify the countries and their contributions in generating tweets and re-tweets. The top 5 countries re-tweeting are Japan, Spain, Brazil, United Kingdom, and the United States whereas in generating Tweets, the top 5 are United Kingdom, Venezuela, Japan, Spain and the United States (ascending order). The USA leads the table in both categories.

\begin{figure}
  \includegraphics[width=0.48\textwidth]{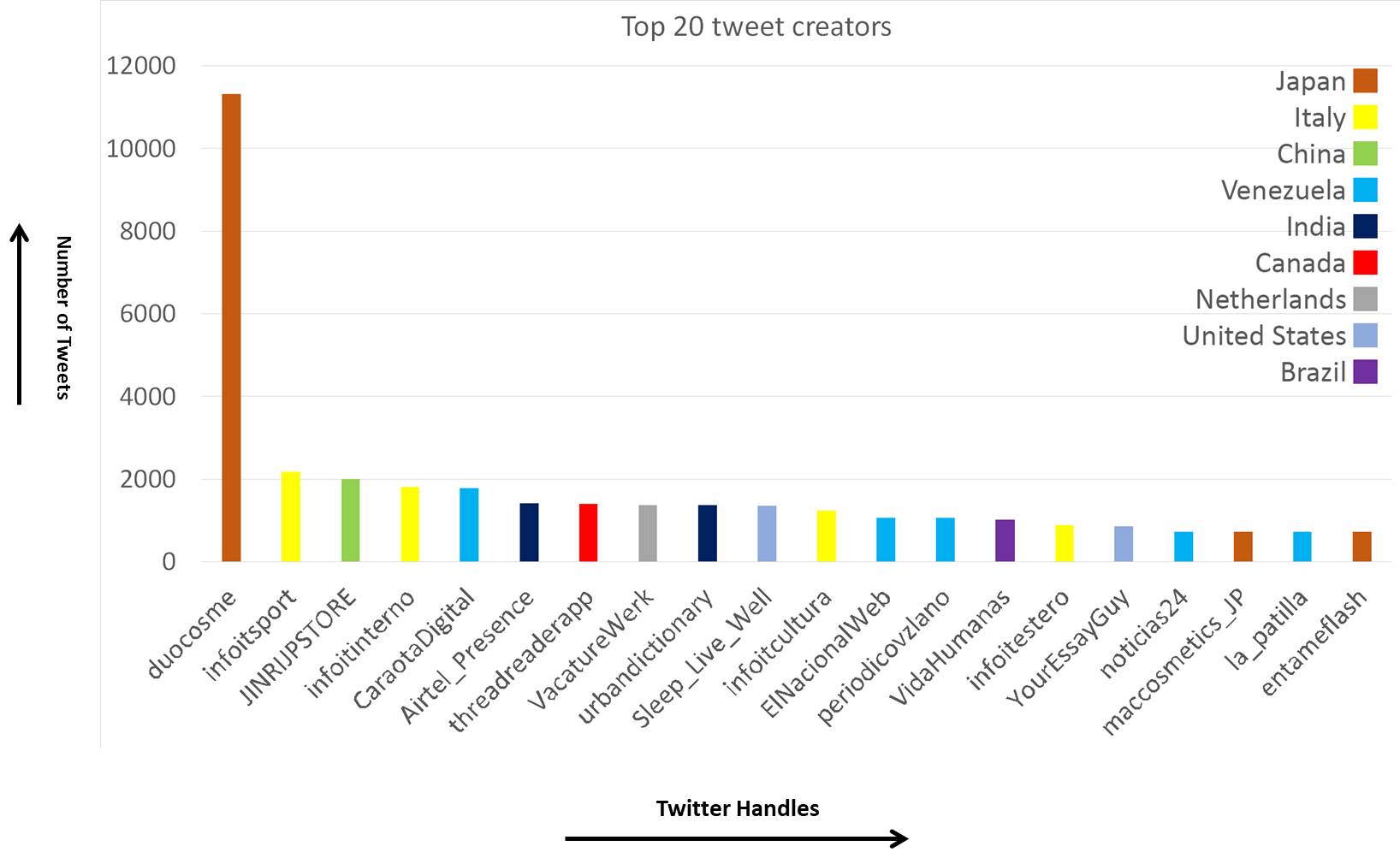}
\caption{Names of top 20 Twitter handles and their country who have identified as most frequent users with respect to the number of tweets during the period of the experiment}
\label{fig:20handles}
\end{figure}

The USA alone contributes 21\% and 37\% in generating Tweets and re-tweets, respectively. Interestingly, the number of distinct countries in generating the top 500 tweets is significantly less than in generating re-tweets. The total countries involved in producing the top 500 twitter handles are 22, however, for generating re-tweets, 50 countries have been reported which implies that the phenomenon of re-tweeting is spread over a larger number of countries. A possible reason may be the fewer efforts and lack of self-innovations and expression of feelings to others.  Moreover, in figure \ref{fig:20handles}, the country producing the top 20 Twitter handles and their names (ID) has been displayed. The colour of the bar represents the countries. We can see from the figure, there are three Twitter handles from Japan in top 20 positions where \textit{‘ducosome’}, a Japan-based company, is the most tweet-creator during the period of experiment with 11,319 tweets. If the number of users is taken into consideration, Venezuela is the top performer with five Twitter handles, whereas Japan is the top performer as far as the number of tweets is concerned.  However, The USA has only two users among the top 20.

Here, we have also investigated the influence of the country's human development index (as released in UN human development report[45]) to the usage of social networking in their region. The details of the countries and their development index is given in the UN report. The UN has classified the countries into four different categories viz. \textit{1) very high human development indices, 2) high human development indices, 3) medium human development indices, and 4) low human development indices}. We have compared the correlation between the rank of these countries as given by the UN and the rank we have assigned on the basis of Tweets generated from the respective countries. Top 20 countries from the first three categories and last 20 from the list of the fourth category are chosen for the comparison. We have analysed the modified spearman rank correlation coefficient for the purpose. Spearman rank correlation is usually calculated for non-parametric values where we may not have two full lists of ranking, i.e. any ranking can be found missing in either of the rankings. Since it is not necessary to have all the rank in both the UN list as well as in the ranking based on tweets for the countries under consideration, we have calculated a modified spearman rank correlation coefficient.

\begin{figure}
  \includegraphics[width=0.48\textwidth]{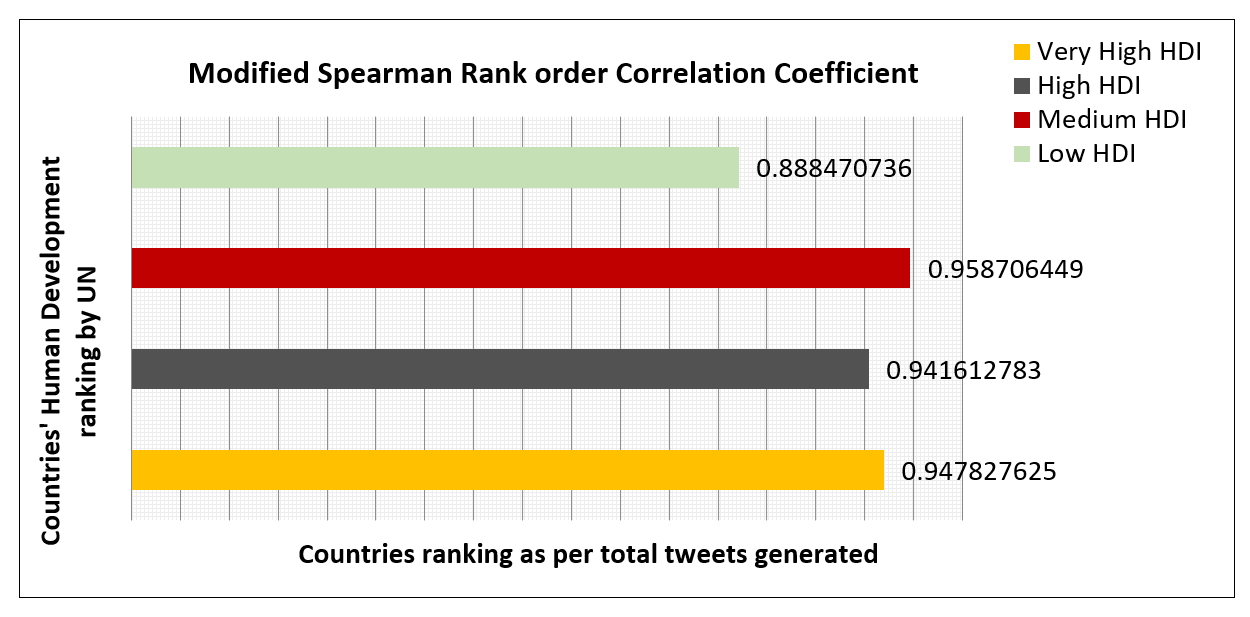}
\caption{Modified Spearman rank correlation coefficient between human development indices (HDI) ranking of the countries given by UN and ranking assigned to these countries on the basis of tweets generated}
\label{fig:correl}
\end{figure}

The correlation values for the countries of all the four categories are very high, the minimum correlation comes out to be 0.8847 for low developed countries and all other have obtained 0.941 or higher values. The graph shows a strong relation between countries human development indices and their tweet trends that indicate that the people from higher developed countries tweet more on twitter and lower developed tweet less. Thus, interestingly we can infer that tweet from a country is highly related to their human development index.

The scatter diagram of the above relation is displayed below from figure \ref{fig:correl} (a) – (d). In sequence, figure \ref{fig:correl} a, b, c and d have a scatter plot between countries ranked according to the tweets generated and low HDI, medium HDI, high HDI, very high HDI, respectively. Countries with human development indices are plotted on Y-axis whereas their corresponding rankings are drawn on X-axis. The scatter diagram shows the close relations between these parameters. 

\begin{figure*}
  \includegraphics[width=1\textwidth]{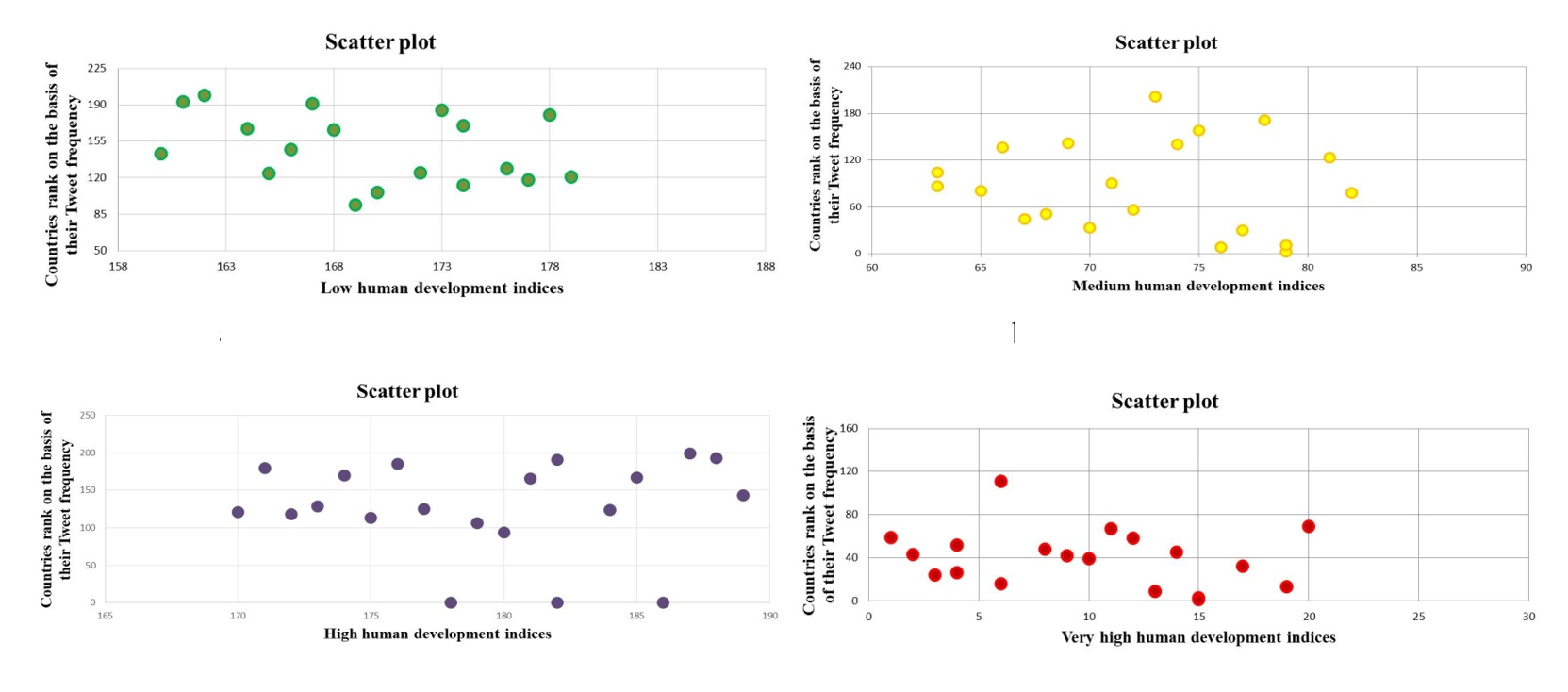}
\caption{Scatter plot between the rank of the countries according to the total tweets and countries ranking by UN development report for a) low human development indices b) medium human development c) high human development indices, and d) very high human development indices}
\label{fig:plot}
\end{figure*}

\begin{figure}
  \includegraphics[width=0.48\textwidth]{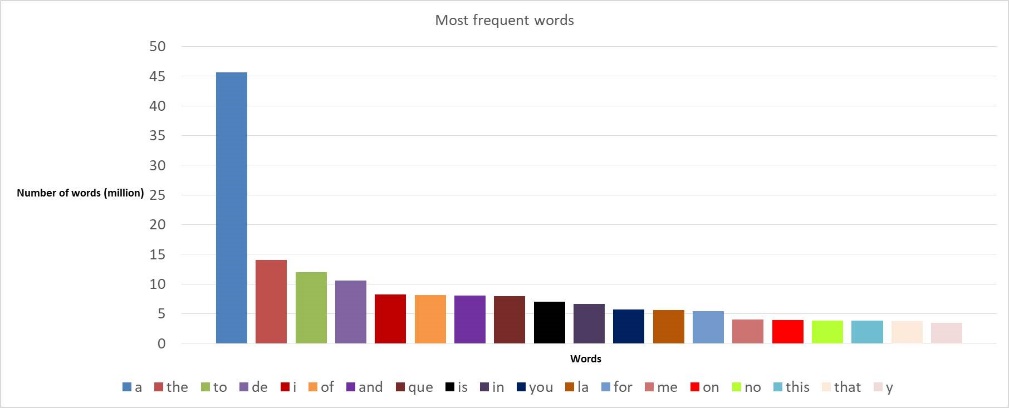}
\caption{Most frequent top 20 words that appear in the tweets worldwide}
\label{fig:words}
\end{figure}

Further, we have tried to explore how users tweet, whether they tweet symbolically, they tweet in English, or the language they use in their daily life. We have considered those countries which are ranked in the top 10 places for their tweet contribution and found out the tweeting behaviour of their users. The detail is given in Table \ref{tab:tweets}. The countries with English as their native language are represented without the row being coloured and the countries with native languages other than English are coloured in green. Canada has two native languages including French along with English, is shown in gray, however, India as a special case with the multi-lingual state has the font in bold, and background coloured in pink. It is evident from the table that people usually tend to tweet in their native language. The percentage of tweets in the native language is much higher which averages 82.61\%. Only the user from India tends to tweet in other than native language, where only 2.5\% tweets were reported in native (national plus regional) languages including Hindi, Bengali, Tamil, Telugu, and Urdu etc. and the reason that English dominates the native languages maybe because of the difference and diversity in Indian languages and culture and its standard of education where English prevails but in daily life, the use of English for a common man is still far away. We have also investigated the most frequent words used in Tweets.  Figure \ref{fig:words} represents the words including articles and prepositions which have been used most frequently in tweets from the users during the period of the experiment. The top words are represented in the figure; \textit{‘a’} is used 45 million times, followed by \textit{‘the’} and \textit{‘to’} which have been used 14 and 12.5 million times, respectively. The other words which have been placed in the ranking are \textit{‘to’, ‘I’, ‘of’, ‘and’, ‘is’, ‘in’, ‘you’, ’for’, ‘me’, ‘on’, ‘no’, ‘this’, ‘that’,} and \textit{‘y’}. The interesting fact is the uses of the non-English words that have scored in the top 20 positions include \textit{‘de’, ‘que’} and \textit{‘la’}. They have got $4^{th}$, $8^{th}$ and $13^{th}$ place in the ranking, respectively. It implies that out of the top 20 words three words are other than English, one from Portuguese and rest are from French languages. However, the word \textit{‘y’} is not a meaningful article or word but it seems an informal way of saying \textit{‘why’} and for others, it is simply to indicate \textit{‘yes’}. When we investigated the above experiment for top 100 words we have obtained, English: 71, Spanish: 21, Portuguese: 3, Dutch: 1, French: 1, and three symbolic representations including \textit{‘y’}, \textit{‘-’} and \textit{‘\%’}. 

\begin{table*}
\caption{Top 10 most tweets contributing countries and the percentage of tweets tweeted in their native language and other languages}
\label{tab:tweets}
\begin{tabular}{lllllll}
\hline\noalign{\smallskip}
Rank & Country & Native Lang. (NL) & Total Tweets & Tweets in NL & \% tweets in NL & Other \% tweets. \\
\noalign{\smallskip}\hline\noalign{\smallskip}
1 & United States & English & 12299716 & 10767530 & 87.5 & 12.5 \\
\rowcolor{green}
2 & Brazil & Portuguese & 3109126 & 2933402 & 94.3 & 5.7 \\
3 & United Kingdom & English & 2732907 & 2660928 & 97.4 & 2.6 \\
\rowcolor{green}
4 & Spain & Spanish & 1630003 & 1325828 & 81.3 & 18.7 \\
\rowcolor{green}
5 & France & French & 1596386 & 1377111 & 86.3 & 13.7 \\
\rowcolor{green}
6 & Argentina & Spanish & 1396407 & 1334201 & 95.5 & 4.5 \\
\rowcolor{pink}
7 & India & Multiple & 1173784 & 28853 & 2.5 & 97.5 \\
\rowcolor{green}
8 & Mexico & Spanish & 1012210 & 874210 & 86.4 & 13.6 \\
\rowcolor{gray}
9 & Canada & English \& French & 800747 & 743336 \& 31285 & 96.7 & 3.3 \\
10 & Nigeria & English & 583463 & 572751 & 98.2 & 1.8 \\
\noalign{\smallskip}\hline
\end{tabular}
\end{table*}

\section{Discussion}
The analysis in this paper gives an insight into the social platform, Twitter. In addition, it lets the reader understand behavioural-trends over Twitter from a geo-graphically-located region of the users, the languages which have been used, the most frequent users and subsequently the countries, and the impact of human development index of the country on the online social platform. 

One of the key findings in this work is related to the location of the users. Although we have considered a total of 49,945,240 tweets for our research, out of which 15,737,327 tweets do not have any location associated with it. In the pre-processing data, it is identified that many users are not sincere while providing their location, and sometimes use fictitious worlds like \textit{Asgard}. One potential reason could be either these users are bogus with fake Twitter IDs and never wish to be identified or they may be computer-generated bots. Hence, this opens up another dimension for the research in the field of social networking.  However, the approach suggested in this paper identifies the location of the users with 68\% accuracy, which again adds value to the literature in exploring the related research. 

Further, the paper suggests some interesting relations between the countries which are more active in using social platforms and which are not. We have used statistical measure \textit{‘modified spearman rank-order correlation coefficient (MSRC)’} to assess how closely the countries are related in terms of their ranking in UN development report 2019 and their ranked position as per the frequency of their tweets. The MSRC performs better when the ranking is compared between partial lists [46–49]. Here, we have selected the top 20 members in the ranking from UN human development report and compared with countries ranking as per the number of tweets. Not all 20 members of the first ranking need to have the place in another ranking from which it is compared; hence, modified spearman rank-order correlation coefficient (MSRC) is given preference over Spearman rank correlation coefficient. Similar is the case with Kendall tau, it does not perform well for a partial list. Therefore, it is worth to use MSRC as a measure to study the correlation between the ranking of countries for UN development report and their ranking as per total tweets they have generated. The countries with very high human development indices tend to generate more tweets and the countries with low human development indices produce lesser tweets. However, India is the only country which is placed by the UN in medium HDI still generates tweets at seventh position. There are the countries in low HDI like Guinea-Bissau and Burkina Faso, from where we could not obtain a single tweet during the period of crawling the data. The above study that gives the relations between countries, their development indices and the trends of their twitter users will enable the various online communities to identify the region where they can run online campaigning, crowdfunding and e-marketing, etc. potentially. Also, the most used language over Twitter is English and most of the countries have their tweets in their native languages. It is also observed that only India has more tweets in English other than the country’s native languages, which includes numerous regional languages. The reason for this behavioural-variation is the fact that India is a diverse country where English is usually a language of colleges and Universities but the people speak the local languages they are used to [50]. The information media is generally used by literate people and they are well familiar with English, hence the most Tweets are obtained in English and not in Hindi or other languages. Further, the regional languages also dominate in India and hence Hindi is not a single speaking language throughout the land and merely used in 7-8 states of the country only. Moreover, the tweets are meant to express one’s feeling to rest of the world; a person who used to speak a language which is known to a particular region would like to use English to convey their message via online social media.

\section{Conclusion and Future work}
In this paper, a detailed analysis of Twittersphere over its 12.8 million members and 86.79 million tweets (initially) has been done and the behaviour of the users, the trends among the countries and the correlation between the human development index of the countries and their corresponding ranking according to the frequency of their tweets have been measured and presented. The result reveals that countries with very high human development indices have a greater number of tweets and countries with low human development indices generate lesser tweets. The paper fulfils two purposes, on one hand, it shall envisage many dimensions of research and gives an analytical view of how human behaves over Twitter, one of the most increasing social media platforms, and on another hand, it gives a simple and versatile way of information processing mechanism.

The present work can be very instrumental for various researches in future, for example, the insight analysis in this article regarding the relationship between countries human development indices and the Twitter users of the corresponding country can serve as a base for social scientists to explore the use of the Internet and subsequently online social media in low income and low human development index countries. Further, the present work also explores the privacy and security issues by addressing the identification of users’ locations. If a user’s identity is not revealed it can lead to mischievous behaviour and may hinder the smooth online activities. The improvement in the identification of the location can help in the concerned field.

\end{document}